\documentclass[useAMS,usenatbib]{mn2e}
\usepackage{graphicx}
\title{A tight scaling relation of dark matter in galaxy clusters} 
\author[Chan]{Man Ho Chan \thanks{mhchan@phy.cuhk.edu.hk} \\
Department of Physics, The Chinese University of Hong Kong,
\\  Shatin, New Territories, Hong Kong, China}

\begin{document}
\date{Accepted XXXX, Received XXXX}

\pagerange{\pageref{firstpage}--\pageref{lastpage}} \pubyear{XXXX}
\maketitle
\label{firstpage}

\begin{abstract}
Recent studies in different types of galaxies reveal that the product of 
the central density and the core radius ($\rho_cr_c$) is a constant. 
However, 
some empirical studies involving galaxy clusters suggest that the product 
$\rho_cr_c$ depends weakly on the total dark halo mass. In this article, 
we re-analyse the hot gas data from 106 clusters and obtain a surprisingly 
tight scaling relation: $\rho_c \propto r_c^{-1.46 \pm 0.16}$. This result 
generally 
agrees with the claims that $\rho_cr_c$ is not a constant for all scales 
of structure. Moreover, this relation does not support the 
velocity-dependent cross section of dark matter if the core formation is 
due to the self-interaction of dark matter.
\end{abstract} 

\section{Introduction} 
The dark matter problem is one of the key issues in 
modern astrophysics. The existence of cold dark matter (CDM) particles is 
the generally accepted model to tackle the dark matter problem. N-body 
simulations show that the dark matter density should follow a universal 
density profile (NFW profile), which goes like $r^{-\alpha}$ towards the 
center of the structure with $\alpha \sim 1-1.5$ \citep{Navarro,Moore}. 
However, observations in many dwarf galaxies and a few 
clusters indicate that flat cores of dark matter exist in those 
structures \citep{Tyson,Gentile,Sand,deBlok,Newman}. This discrepancy can 
be reconciled in many possible scenarios. 
For example, the feedback from baryonic processes such as supernova 
explosion can generate core-like structure in dark matter profile 
\citep{Weinberg,Maccio}. However, some 
studies point out that these processes cannot produce enough feedback to 
get the observed core size \citep{deBlok,Penarrubia,Vogelsberger}.  

Another suggestion is that dark matter particles are 
not collisionless, but weakly self-interacting. Burkert(2000) showed that 
a core could be produced if the dark matter cross-section per unit mass is 
about $\sigma/m \sim 1$ cm$^2$ g$^{-1}$. The resulting dark matter profile 
is known as the Burkert profile. Moreover, recent studies in a wide range 
of galaxies (including dwarf galaxies) report an interesting relation 
if the dark matter density profile is fitted with a cored density profile: 
$\rho_cr_c=$ constant, 
where $\rho_c$ and $r_c$ are the central density and core radius of the 
dark matter density profile respectively. This relation is first noticed 
by Kormendy and Freeman(2004). They obtained $\rho_cr_c \sim 100M_{\odot}$ 
pc$^{-2}$, which is almost a constant by 
using the data from 55 rotation curves in spiral galaxies 
\citep{Kormendy}. 
Later, Spano et al.(2008) analysed the mass distribution of 36 spiral 
galaxies and got 
$\rho_c \propto r_c^{-1.04}$. Furthermore, Donato et al.(2009) use the 
data from 1000 spiral galaxies and obtained $\rho_cr_c 
\approx 141M_{\odot}$ pc$^{-2}$ for a wide range of different galaxies. 
Gentile et al.(2009) also show that this interesting relation can be 
applied in baryonic component of galaxies. Recently, Salucci et al.(2012) 
use the kinematic surveys of the dwarf
spheroidal satellites of the Milky Way to tighten the relation $\rho_c 
\propto r_c^{-a}$ with $0.9<a<1.1$. However, all the above results are 
only based on 
the observations from galaxies. Some studies including the data 
from galaxy clusters suggest that the   
product $\rho_cr_c$ is not really a constant, but depends weakly on the 
total mass of dark matter halo $M_{\rm halo}$. Boyarsky et al.(2009) and 
Del Popolo et al.(2013)
obtained $\rho_cr_c \propto M_{\rm halo}^{0.21}$ and $\rho_cr_c \propto
M_{\rm halo}^{0.16}$ respectively by extending the sample data to cluster 
scale. Moreover, recent observation from cluster 
Abell 611 reports a very large $\rho_cr_c \sim (2350 \pm 200)M_{\odot}$ 
pc$^{-2}$ \citep{Hartwick}. Although the data from clusters are still not 
enough to 
draw any conclusions, these results begin to challenge the universality of 
dark column density ($\rho_cr_c=$ constant).

In fact, the potential relation between $\rho_c$ and $r_c$ suggests that 
some strong constraints or
intrinsic properties may exist in dark matter. Chan(2013a) suggests 
that the 
existence of a universal `optical depth' $\tau= \rho_c(\sigma/m)r_c=$ 
constant in galaxies can explain the observed relation. However, there is 
no strong fundamental reason or physical principle why there exists a 
universal `optical depth'. Therefore, it would be very useful to 
understand the properties of dark matter if we could examine whether the 
universality of dark matter column density is also true in 
galaxy clusters. 

However, the mass density profile probed by gravitational lensing 
cannot provide accurate core radius and central density of dark matter 
profile. Although the method of weak and 
strong lensing can give a good direct estimation of projected mass, the 
3-D mass function still depends strongly on the dark matter functional 
form, 
which is usually assumed to be the NFW profile or generalized NFW profile
\citep{Bartelmann,Mahdavi,Giocoli}. However, as mentioned above, the NFW 
profile deviates from the observed profile significantly for small radius 
and an additional free parameter is needed to indicate the inner slope of 
the density profile \citep{Giocoli}. Moreover, the generalized NFW profile 
can give us the density scale only, but not the central density for our 
analysis. Although \citet{Bartelmann,Shan} are able to obtain a large 
sample of enclosed cluster mass by 
strong lensing, we still need to assume some cored-profile (with $\rho_c$ 
and $r_c$) to de-project the enclosed mass for our purpose. The result, 
however, would be highly dependent on the assumed profile.
  
Alternatively, observations in cluster hot gas provide a good tool to 
probe the dark matter density profile. Although we need to 
assume that the hot gas particles are in hydrostatic equilibrium and the 
distribution is spherically symmetric, we need not assume any 
cored-profile in the analysis to get the $\rho_c$ and $r_c$. 
Since the dark matter mass dominates the total mass, the resulting profile 
can be regarded as the dark matter mass profile. However, the 
spherical asymmetry, cooling flow in hot gas and the AGN feedbacks may 
significantly affect the estimated profile. In particular, the cooling 
flow and AGN feedbacks mainly affect the central part of the 
density profile in clusters by a factor of 2-4 
\citep{Arabadjis,Martizzi}. Although these effects are not negligible, it 
is the only way to obtain a large sample of clusters with corresponding 
$\rho_c$ and $r_c$. Moreover, we can divide the analysis into subsets 
such that the cooling-flow clusters can be ruled out in the empirical 
fits. In this article, 
we still use this traditional method to get a universal dark matter 
density profile for 
106 galaxy clusters from observations based on the ROSAT All-Sky Survey 
\citep{Reiprich}. By relating the central densities and the core 
radii of these density profiles, we can test the universality of dark 
column density for the whole sample and its subsets of clusters.

\section{Mass profile in clusters}
The hot gas density profile can be modelled by the King's profile 
(single-$\beta$ model) \citep{King}
\begin{equation}
n(r)=n_0 \left(1+ \frac{r^2}{r_0^2} \right)^{-3 \beta/2},
\end{equation}
where $n_0$, $r_0$ and $\beta$ are the fitting parameters. Assuming the 
hot gas is in hydrostatic equilibrium and spherically symmetric, we get
\begin{equation}
\frac{kT}{m_g} \frac{dn(r)}{dr}=- \frac{GM(r)n(r)}{r^2},
\end{equation}
where $m_g$ is the average mass of a hot gas particle, $T$ is the hot gas 
temperature and $M(r)$ is the enclosed mass profile in a cluster. By 
combining the Eq.~(1) and (2), we get \citep{Reiprich}
\begin{equation}
M(r)= \frac{3 \beta kTr^3}{Gm_g(r^2+r_0^2)}.
\end{equation}
Here we have assumed that the temperature of hot gas is constant. Although 
the isothermal profile is not a good assumption for many clusters 
\citep{Vikhlinin}, the 
estimation of cluster total mass by using Eq.~(3) is still a good 
approximation for our 
interested region: $r \leq r_0$ \citep{Allen,Reiprich}. It can be 
justified by assuming an approximated form of hot gas temperature 
profile \citep{Pointecouteau}:
\begin{equation}
T(r)=T_0+T_1 \left[ \frac{(r/r_c')^{\eta}}{1+(r/r_c')^{\eta}} \right],
\end{equation}
where $T_0$, $T_1$, $r_c'$ and $\eta$ are fitting parameters. If we 
include the temperature variation, the mass and density profile would be 
respectively modified to
\begin{equation}
M(r)=\frac{kTr}{Gm_g} \left( \frac{3 \beta x^2}{1+x^2}-a \right)
\end{equation}
and
\begin{equation}
\rho(r)=\frac{kT}{4 \pi Gm_gr^2} 
\left[ \frac{3 \beta x^2(3+x^2)(1+a)}{(1+x^2)^2}-a \left(1+\eta- \frac{2 
\eta y^{\eta}}{1+y^{\eta}} \right) \right],
\end{equation}
where $a=d \ln T/d \ln r$, $x=r/r_0$ and $y=r/r_c'$. The term $a$ is 
small for some nearly isothermal clusters. However, for some cool-cored 
clusters, the temperature variation near the centre is not negligible. 
Although this value is not universal for all clusters, \citet{Hudson} 
analysed a large sample of clusters and obtained the average values of 
$a$ for 
cool-cored clusters ($a=0.24$) and non-cool-cored clusters ($a=0.08$) near 
the centres of clusters. By using Eq.~6, the term $a$ would contribute 3\% 
and 25\% errors in the estimation of $\rho_c$ respectively for 
non-cool-cored clusters and cool-cored clusters. These errors are 
generally small compared 
with the observational errors of the required parameters used in the 
calculation. For the outer region, although we do not have a full set of 
fitting parameters for 106 clusters, 
we can extrapolate the result obtained from \citet{Allen}, which got a 
narrow range of fitting parameters for several clusters: 
$T_0=(0.40 \pm 0.02)T$, $T_1=(0.61 \pm 0.07)T$, $r_c'=(0.087 \pm 
0.011)r_{2500}$ and $\eta = 1.9 \pm 0.4$. The virial radius $r_{2500}$ is 
related with another virial radius $r_{500}$ by $r_{2500} \approx 
0.45r_{500}$ for normal clusters. By using the analysis of $r_0$ and 
virial mass of clusters $M_{500}$ from \citet{Chen}, we can obtain a 
relation $r_0 \propto r_{2500}^{3.54}$. Therefore, we get $y \approx 2.5$ 
when $r=r_0$ 
for normal clusters. By putting this estimation to Eq.~(6), the percentage 
error of $r_c$ is just 6 \%, which is much smaller than the observational 
errors of $T$, $\beta$ and $r_0$ (the definition of $r_c$ will be 
discussed below). The non-isothermal factor would be large 
if we consider the total mass 
at large radius. Therefore, for simplicity, we use Eq.~(3) to model all 
the cluster total mass.

Moreover, some hot gas profiles in clusters cannot be well fitted by 
the single-$\beta$ model. \citet{Chen} re-analysed the data from 
\citet{Reiprich} and discovered that the empirical fits for 49 cluster hot 
gas profiles can be significantly improved by using a double-$\beta$ 
model:
\begin{equation}
n(r)=\sqrt{n_{01}^2 \left(1+ \frac{r^2}{r_{01}^2} \right)^{-3 
\beta_1}+n_{02}^2 \left(1+ 
\frac{r^2}{r_{02}^2} \right)^{-3 \beta_2}},
\end{equation}
where $n_{01}$, $n_{02}$, $r_{01}$, $r_{02}$, $\beta_1$ and $\beta_2$ are 
the fitting parameters. By using Eq.~(2), the total mass profile becomes
\begin{equation}
M(r)= \frac{3kTr^3}{Gm_g} \frac{n_{01}^2 \beta_1 
\alpha_1^{-3\beta_1-1}/r_{01}^2+n_{02}^2 \beta_2 \alpha_2^{-3 
\beta_2-1}/r_{02}^2}{n_{01}^2 \alpha_1^{-3 \beta_1}+n_{02}^2 \alpha_2^{-3 
\beta_2}},
\end{equation}
where $\alpha_1=1+(r/r_{01})^2$ and $\alpha_2=1+(r/r_{02})^2$. 

By using Eqs.~(3) and (8), the central mass density of the dark 
matter for single-$\beta$ model and double-$\beta$ model can be 
respectively given by
\begin{equation}
\rho_c= \frac{9 \beta kT}{4 \pi Gm_gr_0^2},
\end{equation}
and
\begin{equation}
\rho_c= \frac{9kT}{4 \pi Gm_g} \left[ \frac{n_{01}^2 
\beta_1/r_{01}^2+n_{02}^2 \beta_2/r_{02}^2}{n_{01}^2+n_{02}^2} \right].
\end{equation}

The dark matter core radius $r_c$ can be regarded as the scale-length of 
dark matter. It can be defined at which the local dark matter volume 
density reaches a quarter of its central value 
\citep{Burkert,Gentile2,DelPopolo}. We can obtain all the mass density 
profiles $\rho(r)$ for clusters from Eqs.~(3) and (8) and follow the usual 
definition to set $\rho(r_c)=\rho_c/4$ to obtain all values of $r_c$.

\section{Data analysis}
A large sample of clusters have been examined by ROSAT All-Sky Survey, 
including the observed parameters $\beta$, $T$ and $r_0$ \citep{Reiprich}. 
They obtained some useful relations such as the correlation between 
the 
luminosity and total mass of clusters ($L-M$ relation). Later, \citet{Ota} 
and \citet{Chen} use an improved sample to obtain some other relations, 
such as $\beta-r_c$, $T-r_c$, $M-T$ and $L-M$ relations. However, they 
didn't analyse the $\rho_c-r_c$ relation. In Fig.~1, we simply 
take the improved sample from \citet{Chen} and plot $\log \rho_c$ vs $\log 
r_c$. A tight relation between $\rho_c$ and $r_c$ is obtained. By fitting 
all cluster hot gas with the single-$\beta$ model, the BCES 
bisector analysis obtains 
\begin{equation}
\log \left( \frac{\rho_c}{M_{\odot}{\rm pc^{-3}}} \right)=(-1.47 \pm 0.04) 
\log \left( \frac{r_c}{1~ \rm kpc} \right)+(0.75 \pm 0.08).
\end{equation}
Since a better fits can be obtained by the double-$\beta$ model for 49 
clusters \citep{Chen}, the same analysis with this improved sample gives
\begin{equation}
\log \left( \frac{\rho_c}{M_{\odot}{\rm pc^{-3}}} \right)=(-1.46 \pm 0.16) 
\log \left( \frac{r_c}{1~ \rm kpc} \right)+(0.88 \pm 0.33).
\end{equation}
A larger uncertainty is resulted in the improved sample because the 
percentage 
error in $r_c$ is generally larger in the double-$\beta$ model, though the 
$\chi^2$ value is smaller \citep{Chen}. Moreover, in Fig.~1, we compare 
the scaling relation in galaxies $\rho_c \propto r_c^{-1}$ obtained by 
\citet{Salucci}. Obviously, the data 
from galaxies and clusters scatter in different parameter space. 
Therefore, they should correspond to different scaling relations. This 
suggests that the product $\rho_cr_c$ is not a universal constant for 
different scales of structure. 

Since cooling flow may affect the mass profile in clusters, we also 
obtain the $\rho_c-r_c$ relation for cooling flow clusters and 
non-cooling flow clusters separately in Fig.~2. The BCES bisector analysis 
gives
\begin{equation} 
\log \left( \frac{\rho_c}{M_{\odot}{\rm pc^{-3}}} \right)=(-1.30 \pm 0.07) 
\log \left( \frac{r_c}{1~ \rm kpc} \right)+(0.60 \pm 0.11)
\end{equation}
for cooling flow clusters and
\begin{equation}
\log \left( \frac{\rho_c}{M_{\odot}{\rm pc^{-3}}} \right)=(-1.50 \pm 0.24) 
\log \left( \frac{r_c}{1~ \rm kpc} \right)+(0.96 \pm 0.54)
\end{equation}
for non-cooling flow clusters. Obviously, the fitted slope for 
non-cooling flow clusters is very closed to the whole sample.

We also perform a cross-check with another sample from \citet{Shan}, 
which 
studied 27 clusters by combining the X-ray and lensing results. The BCES 
bisector analysis obtains
\begin{equation}
\log \left( \frac{\rho_c}{M_{\odot}{\rm pc^{-3}}} \right)=(-1.64 \pm 0.10) 
\log \left( \frac{r_c}{1~ \rm kpc} \right)+(1.58 \pm 0.21).
\end{equation}
This result generally agrees with our analysis based on 106 clusters 
(see Fig.~3). The larger slope obtained from \citet{Shan}'s sample may be 
due to the 
assumption of the single-$\beta$ model used, which has a generally larger 
$r_c$. Moreover, the average hot gas temperature is larger in 
the sample from \citet{Shan} (average $T=8.3$ keV) than the sample from 
\citet{Chen} (average $T=4.8$ keV). This may also affect the slope in the 
empirical fits.

Besides, the mean value of $\rho_cr_c$ for clusters 
is $\sim 
2000M_{\odot}$ pc$^{-2}$, which generally agrees with the result from 
\citet{Hartwick}, which obtained $\rho_cr_c \sim 2350 \pm 200M_{\odot}$ 
pc$^{-2}$.

\begin{figure*}
\vskip5mm
 \includegraphics[width=84mm]{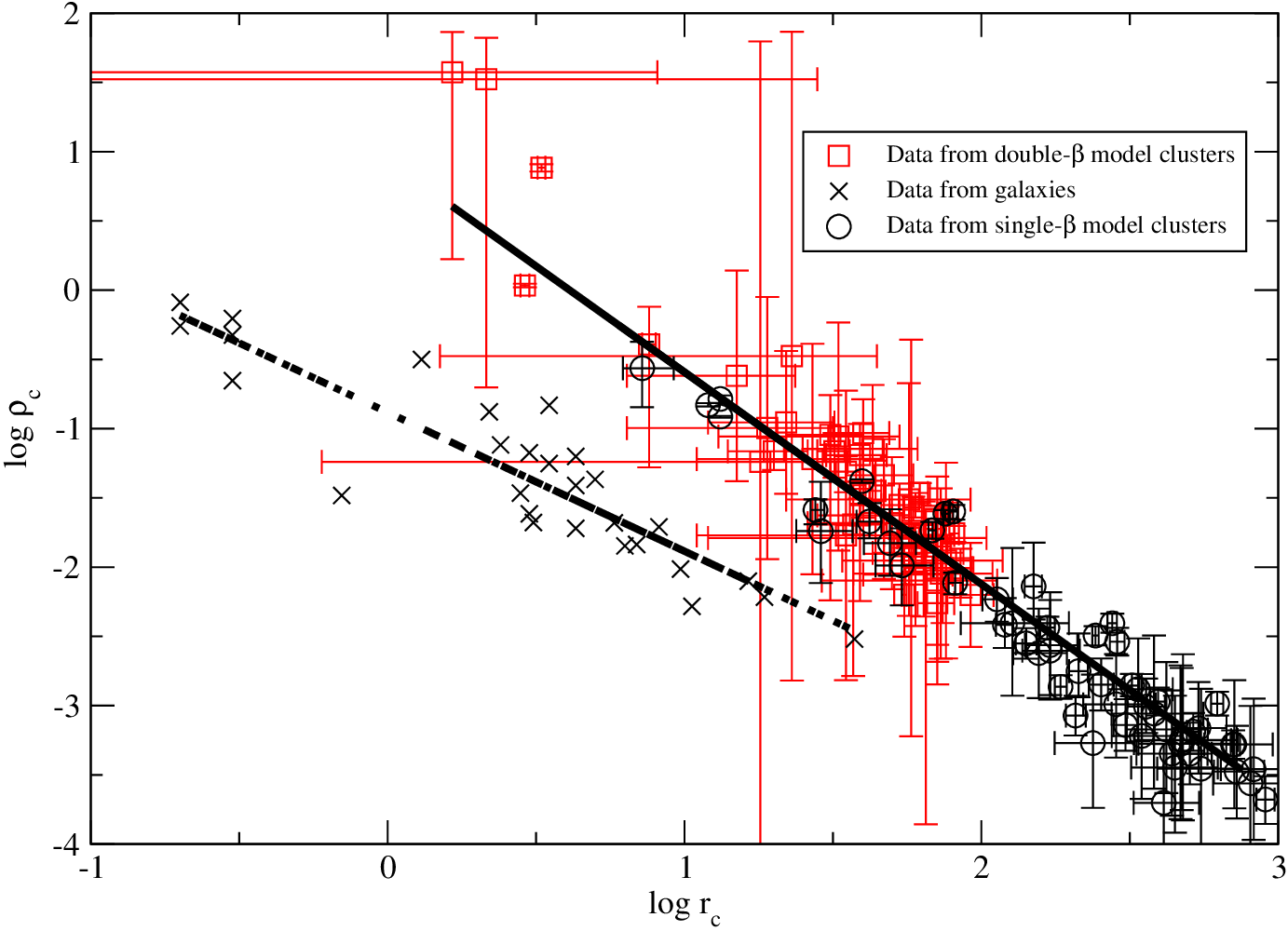}
 \caption{$\log \rho_c$ versus $\log r_c$ from cluster data 
\citep{Chen}. The units for 
$\rho_c$ and $r_c$ are in $M_{\odot}$ pc$^{-3}$ and kpc respectively. The 
fitted slope and the y-intercept for cluster data are $-1.46 \pm 0.16$ 
and $0.88 \pm 0.33$ respectively.}   
\end{figure*}

\begin{figure*}
\vskip5mm
 \includegraphics[width=84mm]{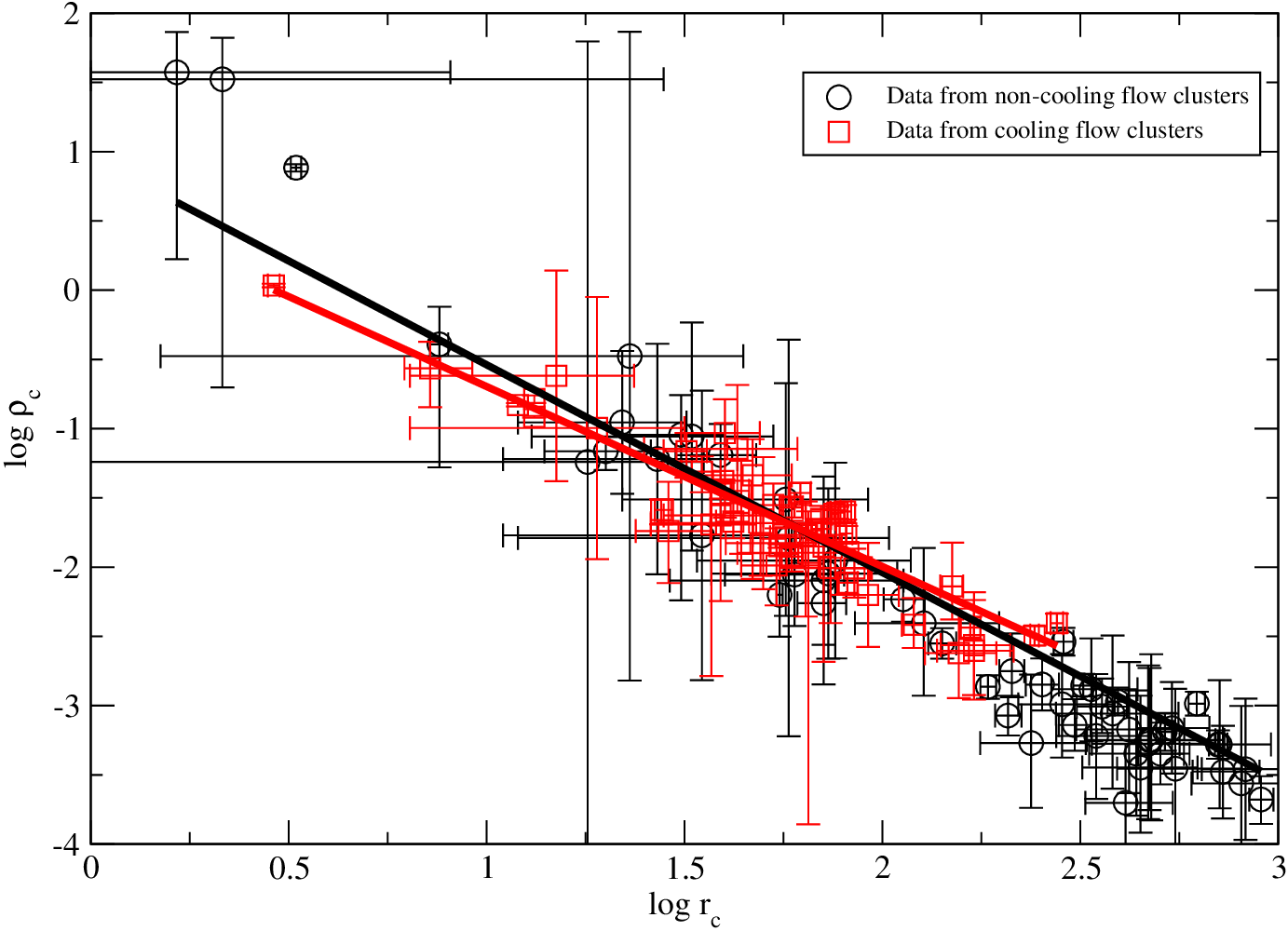}
 \caption{$\log \rho_c$ versus $\log r_c$ from cooling flow cluster data 
(squares) and non-cooling cluster data (circles). The units 
of $\rho_c$ and $r_c$ are in $M_{\odot}$ pc$^{-3}$ and kpc respectively.}   
\end{figure*}

\begin{figure*}
\vskip5mm
 \includegraphics[width=84mm]{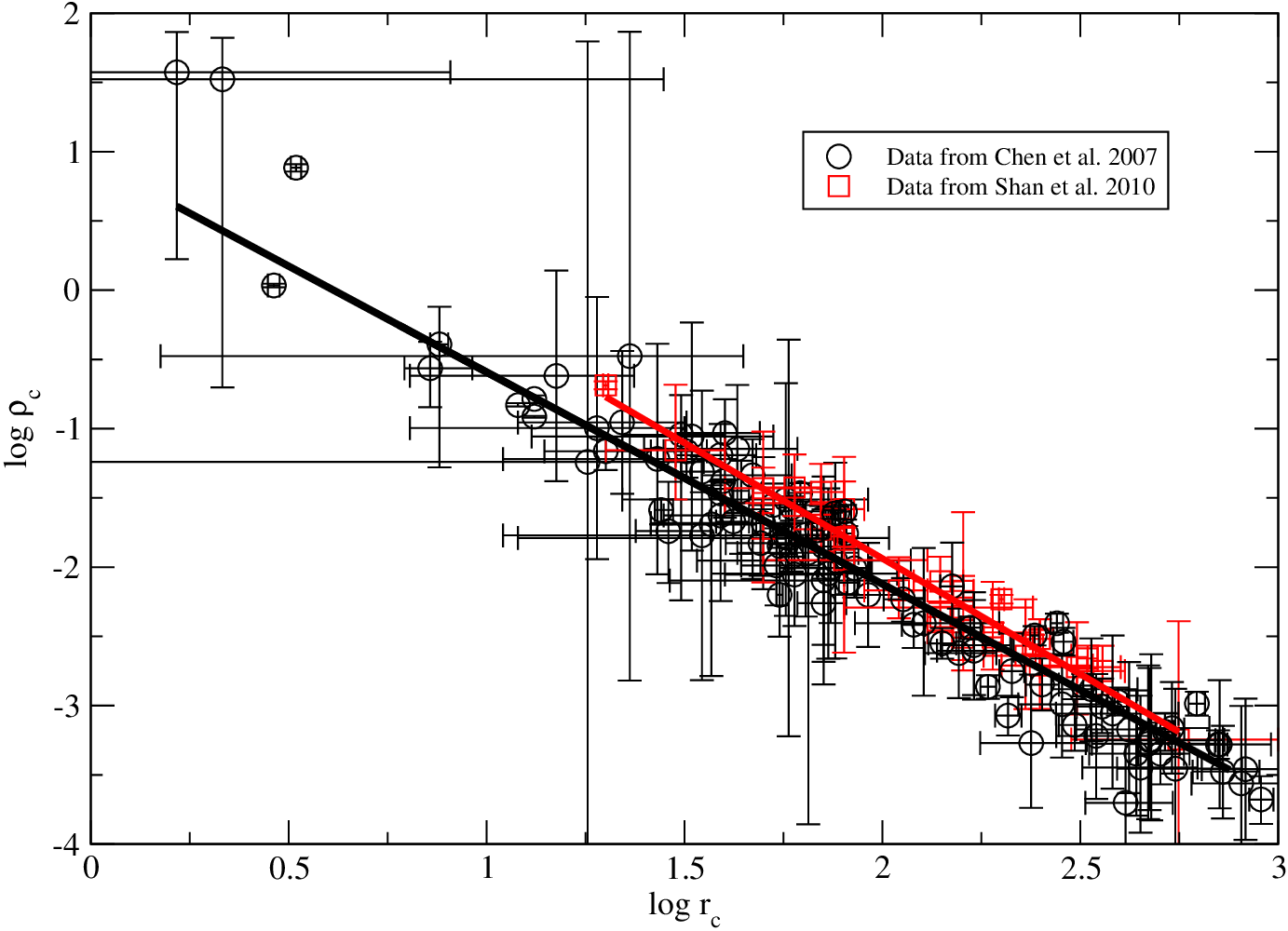}
 \caption{$\log \rho_c$ versus $\log r_c$ from Chen et al. 2007 data 
(circles) and Shan et al. 2010 data (squares). The units 
of $\rho_c$ and $r_c$ are in $M_{\odot}$ pc$^{-3}$ and kpc respectively.}   
\end{figure*}

\section{Discussion and Conclusion}
In this article, we obtain the central density and core radius of 
dark matter in a cluster by using the hot gas profile. The resulting 
scaling relation is $\rho_c \propto r_c^{-1.46 \pm 0.16}$. This result is 
basically different from that obtained in galaxies: $\rho_c \propto 
r_c^{-1}$. Also, this result gives a tighter scaling relation in clusters 
compared with the fits from previous studies such as $\beta \propto 
r_c^{0.11^{+0.03}_{-0.02}}$ and 
$T \propto r_c^{0.03^{+0.05}_{-0.07}}$ \citep{Ota}. On the other hand, the 
fitted slope for cooling flow 
clusters (slope$=-1.30 \pm 0.07$) are slightly
different from non-cooling flow clusters (slope$=-1.50 \pm 0.24$). This 
suggests that the cooling flow in clusters may affect the inner structure 
of dark matter, which is consistent with some suggestions about the 
baryonic feedbacks in galaxies \citep{Weinberg,Maccio}.

The scaling relation in clusters consists of three basic 
parameters in hot gas profile, $\beta$, $r_0$ and $T$. These parameters 
should be independent of each other in hot gas. However, the 
gravitational interaction between dark matter and hot gas particles 
relates the three parameters to form a tight scaling relation. Therefore, 
this scaling relation may reflect some intrinsic properties of dark matter 
in clusters. If the core formation is due to the self-interaction of dark 
matter, as suggested by Spergel and Steinhardt(2000), the scattering cross 
section 
$\sigma$ should be related to $(\rho_cr_c)^{-1}$. In galactic scale, since 
$\rho_cr_c$ is a constant, $\sigma$ is also a constant for galaxies. This 
supports the constant (velocity independent) self-interaction cross 
section scenario \citep{Peter,Rocha,Zavala}. However, recent studies in 
clusters have already revealed that $\sigma$ should be velocity-dependent 
($\sigma$ decreases as the velocity of dark matter particle increases) 
\citep{Loeb,Chan2}. By using our result from cluster data, since 
$\rho_cr_c^{1.46}$ is a constant, we have $\sigma \sim (\rho_cr_c)^{-1} 
\propto r_c^{0.46}$. Generally, the velocity of dark matter particle $v$ 
increases with $r_c$ \citep{Rocha}. That means $\sigma$ should increase 
with $v$, which contradicts to the observations and the prediction from 
velocity-dependent cross section scenario \citep{Colin,Vogelsberger}. 
Therefore, the formation of cores in clusters and galaxies 
may not be caused by the self-interaction of dark matter particles.     

Finally, the two very different scaling relations in galaxies and 
clusters suggest that some different constraints in 
dark matter may exist in galaxies and clusters. Therefore, probably there 
will be no universal dark matter density profile exist in different 
scales as predicted by numerical simulations.     

\section{Acknowledgements}
I am grateful to the referee for helpful comments on the manuscript.

\label{lastpage}
\end{document}